\documentclass[letterpaper]{article} 
\usepackage{aaai25}  
\usepackage{times}  
\usepackage{helvet}  
\usepackage{courier}  
\usepackage[hyphens]{url}  
\usepackage{graphicx} 
\usepackage{natbib}  
\usepackage{caption} 
\usepackage{algorithm}
\usepackage{algorithmic}

\usepackage{array}
\usepackage{multirow}
\usepackage{booktabs}
\usepackage{amsmath}
\usepackage{bbding}
\usepackage{enumitem}\newcommand{\modelname}{EyEar}
\usepackage{newfloat}
\usepackage{listings}
\DeclareCaptionStyle{ruled}{labelfont=normalfont,labelsep=colon,strut=off} 
\floatstyle{ruled}
\newfloat{listing}{tb}{lst}{}
\floatname{listing}{Listing}
\title{EyEar: Learning Audio Synchronized Human Gaze Trajectory \\ Based on Physics-Informed Dynamics}
\author{
    Xiaochuan Liu\equalcontrib\textsuperscript{\rm 1},
    Xin Cheng\equalcontrib\textsuperscript{\rm 1},
    Yuchong Sun\textsuperscript{\rm 1},
    Xiaoxue Wu\textsuperscript{\rm 1},
    Ruihua Song\textsuperscript{\rm 1}\thanks{Corresponding authors.},
    Hao Sun\textsuperscript{\rm 1}\footnotemark[2],\\
    Denghao Zhang\textsuperscript{\rm 2}\footnotemark[2]
}
\affiliations{
    \textsuperscript{\rm 1}Gaoling School of Artifcial Intelligence, Renmin University of China, Beijing, China\\
    \textsuperscript{\rm 2}Department of Psychology, Renmin University of China, Beijing, China\\
    \{liuxiaochuan, chengxin000, ycsun, wuxiaoxue1102, rsong, haosun, zdh\}@ruc.edu.cn
    
}

\usepackage{bibentry}

\begin{document}

\maketitle

\begin{abstract}
Imitating how humans move their gaze in a visual scene is a vital research problem for both visual understanding and psychology, kindling crucial applications such as building alive virtual characters. Previous studies aim to predict gaze trajectories when humans are free-viewing an image, searching for required targets, or looking for clues to answer questions in an image. While these tasks focus on visual-centric scenarios, humans move their gaze also along with audio signal inputs in more common scenarios. To fill this gap, we introduce a new task that predicts human gaze trajectories in a visual scene with synchronized audio inputs and provide a new dataset containing 20k gaze points from 8 subjects. To effectively integrate audio information and simulate the dynamic process of human gaze motion, we propose a novel learning framework called \modelname~(Eye moving while Ear listening) based on physics-informed dynamics, which considers three key factors to predict gazes: eye inherent motion tendency, vision salient attraction, and audio semantic attraction. We also propose a probability density score to overcome the high individual variability of gaze trajectories, thereby improving the stabilization of optimization and the reliability of the evaluation. Experimental results show that \modelname~outperforms all the baselines in the context of all evaluation metrics, thanks to the proposed components in the learning model.
\end{abstract}

%

\section{Introduction}
\begin{figure}[t]
  \centering
  \includegraphics[width=\linewidth]{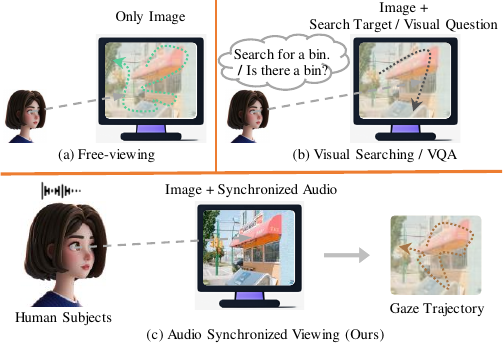}
  \caption{Different from existing tasks which predict gazes in situations such as (a) humans free-viewing an image, or (b) searching for required targets or clues to answer questions in an image,
  our proposed task (c) aims to predict human gaze in a more common scenario where humans receive synchronized audio signals when directing their gaze.} 
  \label{example}
  \vspace{-1em}
\end{figure}

Predicting human gaze trajectory within a visual scene
is an important research problem for a range of research communities, including visual understanding, human behavior, psychology, etc. It is also crucial for some downstream applications, such as building virtual characters that are alive and more interactive.
As shown in Figure~\ref{example}, existing studies of gaze trajectory prediction aim to predict gaze sequences (a.k.a. scanpaths) when human are free-viewing images~\cite{jiang2015-dataset} or webpages~\cite{shen2014-webpagea}, searching for required objects in an image~\cite{mondal2023gazeformer,yang2020-predicting}, or collecting clues for answering visual questions~\cite{chen2021predicting}.

While existing tasks focus on visual-centric scenarios, 
the gaze trajectory of humans can be affected by both visual and audio stimuli.
To bridge this gap, 
we propose a new task: predicting a gaze trajectory in an audio synchronized viewing scene. We contribute to this task from both dataset and algorithm perspectives.

From the dataset perspective, we construct a new dataset with 20k gaze points from 8 subjects, using eye tracking devices to trace the subjects' scanpaths over the image when they hear an audio clip.
Our dataset has a much longer average gaze sequence length and duration than existing datasets, making it more challenging and valuable for learning long-range human gaze trajectories.

From the algorithm perspective, we solve two major challenges. The first challenge is how to effectively integrate audio information in this multimodal scenario, with a proper framework that considers the dynamical physics of eye movement (\emph{C1}). 
The second challenge is that different people have diverse gaze trajectories, making the collected data have high individual variability (\emph{C2}). 
This makes the commonly used Mean Square Error (MSE) loss confused and raises a challenge for the stability of optimization. 
Existing evaluation metrics based on Euclidean distance also lack reliability due to divergent targets.

We overcome the aforementioned challenges by proposing an \modelname~framework with two novel designs.
First, we build an audio-aware dynamical system (\emph{C1}), inspired by the dynamical systems in physics which can model the complex evolution process using states and motion vectors~\cite{birkhoff1927dynamical}.
Specifically, we consider three kinds of forces to decide motion: an inherent motion tendency force to model the continuous transitions of eye movement, an audio semantic attraction force that models fine-grained audio-visual association to predict an attraction point in the image due to the presence of audio, and a vision salient attraction force that models attractions to salient image regions.
Second, we propose a probability density score (PDS) for reliable evaluation and a corresponding probability density loss to ensure stable optimization (\emph{C2}). PDS is based on mixed Gaussian distributions fitted from the gaze points of multiple subjects. It estimates the distribution using Gaussian kernel density estimation and measures how well the predicted gaze point fits the ground-truth distribution by the normalized value of its probability density.

Experiment results show that~\modelname~performs the best in all evaluation metrics. The improvements over the best in each metric are from 4\% to 15\%.
Ablation studies show that our proposed components are all effective for improving performance.
Our contributions can be summarized as follows:

\begin{itemize}[leftmargin=1.2em]
    \item We propose a new task that predicts gaze trajectories in a visual scene under the stimulation of synchronized audio inputs. We also collect a dataset containing 20k gaze points from 8 subjects to facilitate the investigation.
    \item We propose an~\modelname~framework that models physics-informed dynamics by considering three possible forces that influence motion to effectively integrate audio information by fine-grained audio-vision association modeling and simulate the dynamics of human gaze.
    \item We propose a probability density score and loss based on mixed Gaussian distributions of multiple gaze trajectories, which not only stabilize model optimization but also evaluate different methods more reliably.
\end{itemize}

\section{Related Work}
\begin{table*}[!ht]
\begin{center}
    \centering
\begin{tabular}{lcccc}
   \toprule
   \multirow{2}{*}{Dataset} & \multirow{2}{*}{Task Scenario} & Synchronized  & Avg. Sequence & Avg. \\
   &&Stimuli&Length&Duration \\
   \midrule
   OSIE~\cite{xu2014predicting} &  Free-viewing & \XSolidBrush & 9.36 & 2.01s \\ 
   COCO-FreeView~\cite{chen2022characterizing} & Free-viewing & \XSolidBrush & 15.45 & 4.05s \\ 
   COCO-Search18~\cite{chen2021coco} & Visual Searching & \XSolidBrush &  3.77 & 0.92s \\
   AiR~\cite{chen2020air} & Visual Question Answering & \XSolidBrush & 10.16 & 2.89s \\  
   \midrule
   EyEar-20k (Ours) & Audio Synchronized Viewing & \Checkmark & \textbf{26.06} & \textbf{12.09s} \\
   \bottomrule
\end{tabular}
\caption{Comparing our dataset with commonly used gaze trajectory prediction datasets. Different from free-viewing or visual searching, our task predicts gaze trajectory in a visual scene with synchronized audio stimuli. Our dataset has much longer sequence length and duration than existing datasets, enabling learning long-range human gaze dynamics from our dataset.
}
\label{table:dataset}
\end{center}
\end{table*}

\textbf{Gaze Trajectory Prediction.}
Studies in the neuroscience field \cite{moschovakis2001functional, krauzlis2013superior}
show that when staring at a certain position on an image, human brain selects the next point to look at and then moves to it. To mimic this mechanism and generate human-like gaze sequences, researchers propose the gaze trajectory (a.k.a scanpath) prediction tasks. Different gaze trajectory prediction tasks focus on different specific tasks such as free-viewing~\cite{judd2012-benchmark, borji2015-cat2000b, jiang2015-dataset}, viewing webpages~\cite{shen2014-webpagea}, visual searching~\cite{yang2020-predicting, mondal2023gazeformer}, and visual question answering~\cite{chen2021predicting}. 
\citet{kummerer2021-stateoftheart} survey the models that predict gaze trajectories and classify previous models into four categories: biologically inspired models \cite{engbert2015-spatial, zanca2020-gravitational}, 
statistically inspired models \cite{xia2019-predicting, lan2022-eyesyn},
cognitively inspired models \cite{liu2013-semanticallybased, sun2021-visual} and engineered models \cite{assens2017-saltinet, kummerer2022-deepgaze}. 
However, most existing works focus on silent tasks with well-defined requirements. Instead our work extends the gaze trajectory prediction task to a visual scene with synchronized audio inputs, and proposes a physics-informed model to learn a natural gaze trajectory.

\textbf{Visual Grounding.}
Our task is also related to visual grounding task because people are expected to gaze at the semantically relevant regions of images when hearing narrations. Visual grounding task aims at locating objects queried by natural language in an image. It is a multi-modal task that requires understanding of the relationship between language and vision. Many advanced models like MDETR \cite{kamath2021-mdetr}, GLIP \cite{li2022-grounded}, DQ-DETR \cite{liu2022-dqdetr}, and Grounding DINO \cite{liu2023-grounding} utilize the power of object detection techniques and pre-trained multi-modal models to achieve good performance. MITR~\cite{meng2021-connectinga} is a more relevant work to our task in terms of data sources. 
It is trained on Localized Narratives (LN) dataset~\cite{pont-tuset2020-connecting} with bounding boxes derived from mouse traces while annotators describing images. However, there exists substantial divergence between eye tracking data and mouse tracking data~\cite{tavakoli2017-dataset}. Despite being similar to visual grounding tasks, our new task has more challenges involving the large amount of ungroundable words, the unique motion process of gazes, and the high individual variability of gazes from different persons, which will be addressed in this paper.

\section{Task and Dataset}
\subsection{Problem Formulation}
The input to our task consists of an image $V$ and an audio clip $A$. Using speech recognition tools,
we can obtain a sequence of words with their start and end time from the audio:
\begin{equation}
    (w_1,t^s_1,t^e_1),\cdots,(w_i,t^s_i,t^e_i),\cdots,(w_n,t^s_n,t^e_n), 
\end{equation}
where $w_i$ is the i-th word, $t^s_i$ and $t^e_i$ are the start and the end timestamps of word $w_i$ in audio $A$ respectively, $n$ is the number of words in the audio.

In our task, human gaze well aligns with the audio timestamps. Subjects tend to gaze at a point until they hear the next word. Therefore, we predict gaze points at each end time. At the end time $t^e_i$, ground-truth gaze points from $N$ subjects are recorded. Consequently, for the image-audio pair $V$-$A$, the recorded gaze trajectory of the $j_{th}$ subject is:
\begin{equation}
    S_j = \{s_1^j,\cdots, s_n^j\},  \quad(j=1,\cdots,N),
\end{equation}
where $s$ denotes a coordinate on the image. 
The objective of our task is to predict a gaze trajectory $\hat{S} = \{\hat{s}_1, \cdots, \hat{s}_n\}$ that is most similar to those ground-truth gaze trajectories conditioned on image and audio inputs.

\subsection{Dataset Construction}\label{section:data_construction}
To facilitate our work and future research, we collect gaze tajectory data from eight subjects when they view images and simultaneously listen to corresponding audio clips. Our dataset contains a total of 20k gaze points. The detailed steps for constructing the dataset are as follows.
(1) \textbf{Image Selection.}
We select images from Unsplash.com website, known for providing high-quality, visually appealing images that are freely available to the public. To mirror the real world more accurately, we select images that one might encounter in everyday life. We also require the selected images containing multiple objects and rich details, while avoiding close-up shots. All images have a resolution of 1024*1024 pixels.
(2) \textbf{Narrative Audio Designing.}
We design the narrative text using a combination of automatic generation and manual refinement. In our task, text primarily serves to guide the subjects' attention to different areas of the images. Thus it would provide accurate and detailed descriptions of the images. 
For efficiency, we choose LLaVA-1.5~\cite{liu2023-visual} to automatically generate descriptions of the images.
To ensure data quality, annotators are asked to manually modify the generated descriptions to avoid rigid patterns, such as always starting with ``there is'' or ``in the image'', increase the diversity of description expressions, correct mistakes, and supplement descriptions of objects that have not been described.
The prepared narrative text is converted to audio with a TTS tool
and then played to the subjects. A 5-second blank audio is added at the beginning, allowing the subjects to get familiar with the image.
Please note that all text and audio data in our dataset are in Chinese because all subjects participating in the data collection are native Chinese speakers.
(3) \textbf{Subjects Instructions.} It is important to note that the subjects are not given explicit instructions. They are instructed to maintain the most natural state because we aim to predict gaze trajectories that occur in daily life rather than in a specific task scenario. We introduce other details of our dataset collection in Supplementary~A.

\subsection{Dataset Comparison}
We compare our dataset with commonly used gaze trajectory datasets, as shown in Table~\ref{table:dataset}. OSIE and COCO-FreeView only record human gaze trajectories when free-viewing images without stimuli of other modalities. While COCO-Search18 and AiR datasets contain textual information, they present the text before displaying images, the gaze trajectories reflect the searching process over images. In contrast, our dataset presents subjects images with synchronized auditory stimuli, closely mirroring real-world scenarios. Furthermore, the statistics indicate that our dataset features much longer sequence length and duration, making it more effective for learning long-range human gaze trajectories. 

\section{Method}\label{section:method}

\begin{figure*}[t]
  \centering
   \includegraphics[width=0.9\textwidth]{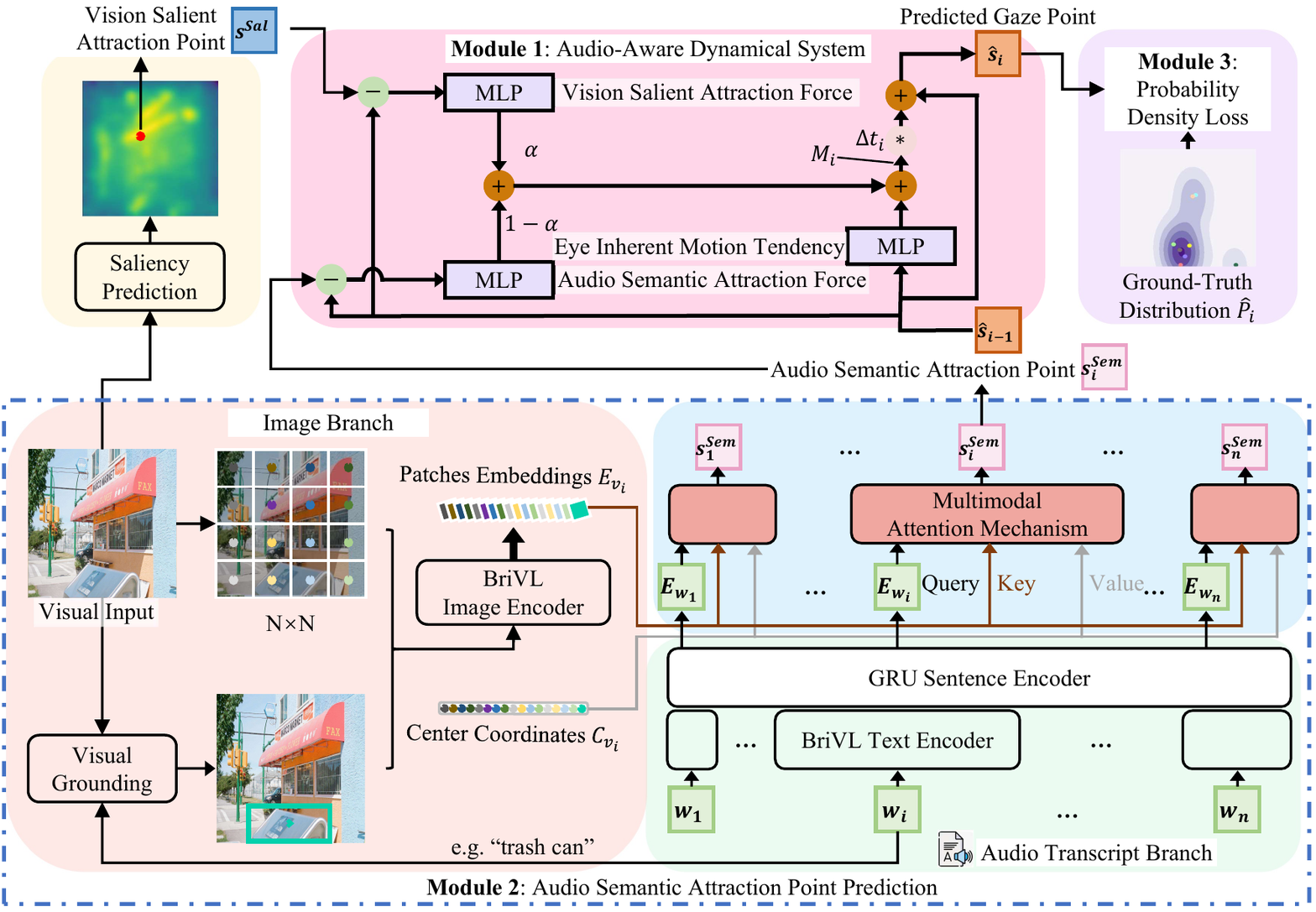} 
  \caption{\textbf{\modelname~overview.} The core component is a physics-informed audio-aware dynamical system that simulates the motion of eyes (See Module 1). The next predicted gaze point is calculated from the current gaze point, the time interval, and a motion vector. The motion vector is influenced by three kinds of forces. The most important force, audio semantic attraction force is predicted by Module 2. We propose probability density loss  (See Module 3) to train the model.}
  \label{overview}
  \vspace{-1em}
\end{figure*}

\subsection{Framework Overview}
As shown in Figure~\ref{overview}, our proposed framework~\modelname~is built on a physics-informed dynamical system that considers three kinds of forces to decide the motion of eyes (See Module 1): one force keeping the inherent motion tendency of eye, one force attracting eyes to salient part of image $V$ no matter what the subject heard, and the other force attracting eyes to the audio semantic attraction point that is semantically relevant to the heard text. 
Specifically, we propose a multimodal attention mechanism to predict the audio semantic attraction point for each word (See Module 2). It can integrate three different types of information: image, text, and image patch coordinates into the coordinates of next audio semantic attraction point. Furthermore,~\modelname~introduces a probability density score function based on mixed Gaussian distributions fitted from multiple ground-truth gaze points as a loss function (See Module 3). It allows for better optimization and evaluation of a natural gaze trajectory, which has high individual variability across humans.

\subsection{Audio-Aware Dynamical System}\label{section:ds}

To capture the motion features of eyes, we propose an audio-aware dynamical system inspired by physics~\cite{birkhoff1927dynamical}. In a dynamical system, there is a concept known as a state, which is a set of determinable real numbers. Tiny variations in the state correspond to tiny variations in these real numbers. The evolution of the dynamical system is governed by a set of functions, describing how future states depend on the current state. The rule is deterministic, which means, for a given time interval, only one future state can evolve from the current state. Herein, a state refers to the gaze location.

As shown in Module 1 of Figure~\ref{overview}, the formulation of our dynamical system is as follows:
\begin{equation}
    \hat{s}_{i} = \hat{s}_{i-1} + \Delta t_{i} \cdot {M
    }_{i},
\label{eq.s1}
\end{equation}
where the current predicted gaze
point $\hat{s}_{i}$ is calculated based on the previous gaze $\hat{s}_{i-1}$, the time interval $\Delta t_{i} = t^e_{i}-t^e_{i-1}$, and a motion vector ${M}_{i}$. Specifically, $M_i$ is calculated as:
\begin{equation}
\begin{split}
    {M}_{i} = \textit{MLP}_A(\hat{s}_{i-1}) &+
    \alpha \cdot \textit{MLP}_B(s^{Sal} - \hat{s}_{i-1}) \\
     &+ (1-\alpha) \cdot\textit{MLP}_C(s^{Sem}_{i} - \hat{s}_{i-1}),
\label{eq.sp}
\end{split}
\end{equation}
where $s^{Sal}$ means vision salient attraction point and $s^{Sem}_i$ means the $i_{th}$ predicted audio semantic attraction point. We comprehensively consider three sources of force influencing the motion vector that represents both the direction and speed of motion. The neural networks in Eq.(\ref{eq.sp}) correspond to the set of functions (motion components caused by three forces) in the dynamical system. The first term in Eq.(\ref{eq.sp}) represents the motion component caused by the force that keeps the inherent motion tendency of eye at the current position $\hat{s}_{i-1}$, regardless of any stimuli.
The second term in Eq.(\ref{eq.sp}) represents the motion component caused by the force that attracts the gaze to the most salient point in the image. $ s^{Sal} $ is obtained by utilizing the DeepGaze IIE~\cite{linardos2021deepgaze} model.
We consider this force because human attention may sometimes be completely captivated by the image, in particular the salient part.
The third term in Eq.(\ref{eq.sp}) represents the motion component caused by the force that attracts the gaze to the audio semantic attraction point $s^{Sem}_{i}$.
This term considers human attention influenced by auditory stimuli. Intuitively, when a human hears some words, she/he pays attention to the semantically relevant part.
Finally, the learnable weighting parameter $ \alpha $ measures the extent to which human attention is drawn to the image itself rather than being influenced by the heard.

\subsection{Audio Semantic Attraction Point Prediction}\label{section:attention}
As shown in Module 2 of Figure~\ref{overview}, to measure the broad semantic relations between image regions and the heard words, we carefully design image branch, audio transcript branch, and a multimodal attention mechanism to integrate different types of information and predict the next audio semantic attraction point.

\textbf{Image Branch.} 
In our scenario, both coarse-grained background information and fine-grained visual grounding information in image $V$ are found helpful in identifying audio semantic attraction points.
For the background information, we segment the image into $N \times N$ patches. We choose $N = 4$ for its best performance in our experiments. We also record the center coordinates of these patches for subsequent use. 
For the fine-grained visual grounding information, we apply the state-of-the-art model Grounding DINO~\cite{liu2023-grounding} to identify the most relevant region in the image for each word $w_i$. This identified region is regarded as a special patch $V_{D_i}$ and we record its center coordinates too. 
When no identified region output by Grounding DINO, we set the embedding of $V_{D_i}$ as all zeros, and the center coordinates as $(0,0)$.
Next, we employ a large-scale pre-trained model to align image and text embedding in the same space. Given the Chinese nature of our narrations, we choose BriVL \cite{huo2021-wenlan} (a Chinese CLIP-like model) for this purpose. All $N \times N$ patches plus the special patch identified by Grounding DINO are finally encoded into embeddings: $ E_{v_i} = \textit{IE}(V_{N \times N},V_{D_i}) $, where IE is the abbreviation of BriVL's Image Encoder.

\textbf{Audio Transcript Branch.} 
We first use the text encoder of BriVL to encode each word in the audio: $ E'_{w_i} = \textit{TE}(w_i) $, where TE is the abbreviation of Text Encoder.
For a large number of 
ungroundable words, the correlation between their embeddings and the image embeddings is minimal. To achieve fine-grained alignment between audio and visual stimuli, we use the GRU for additional sentence encoding: $ E_{w_i}, H_i = \textit{GRU}(H_{i-1}, E'_{w_i}) $, where $H_{i}$ represents the hidden state of the GRU at the $i$-th time step and $E_{w_i}$ represents the final embedding for word $w_i$. We choose the GRU instead of Transformer~\cite{vaswani2017attention} for two reasons: i) we observe that the semantic attraction point is sensitive to the nearer heard words, which well aligns with the characteristics of RNN models~\cite{zaremba2014recurrent}. They can effectively utilize short-range context information; and ii) compared to the Transformer, GRU requires much fewer computational resources and training data.

\textbf{Prediction with Multimodal Attention Mechanism.} 
Intuitively, when a person is viewing an image and listening to narrative audio at the same time, various relevant parts of the image exert attractive forces to her/him. The audio semantic attraction point should be a comprehensive manifestation of these attractive forces. Therefore, we design a multimodal attention mechanism to integrate information from image and text, and finally predict the audio semantic attraction point.
Specifically, we treat text embedding $E_{w_i}$ as query, each image embedding in $E_{v_i}$ as key, and the center coordinates of each image patch as value:
\begin{equation}
    s^{Sem}_{i} = \textit{Attention}(E_{w_i}, E_{v_i}, C_{v_i}),
\end{equation}
where $ C_{v_i} $ represents the center coordinates of patches, and $ s^{Sem}_{i} $ represents the audio semantic attraction point for word $w_i$. The network structure of our attention mechanism is the same as the Transformer block~\cite{vaswani2017attention}.

\begin{figure}[!t]
  \centering
  \includegraphics[width=0.9\linewidth]{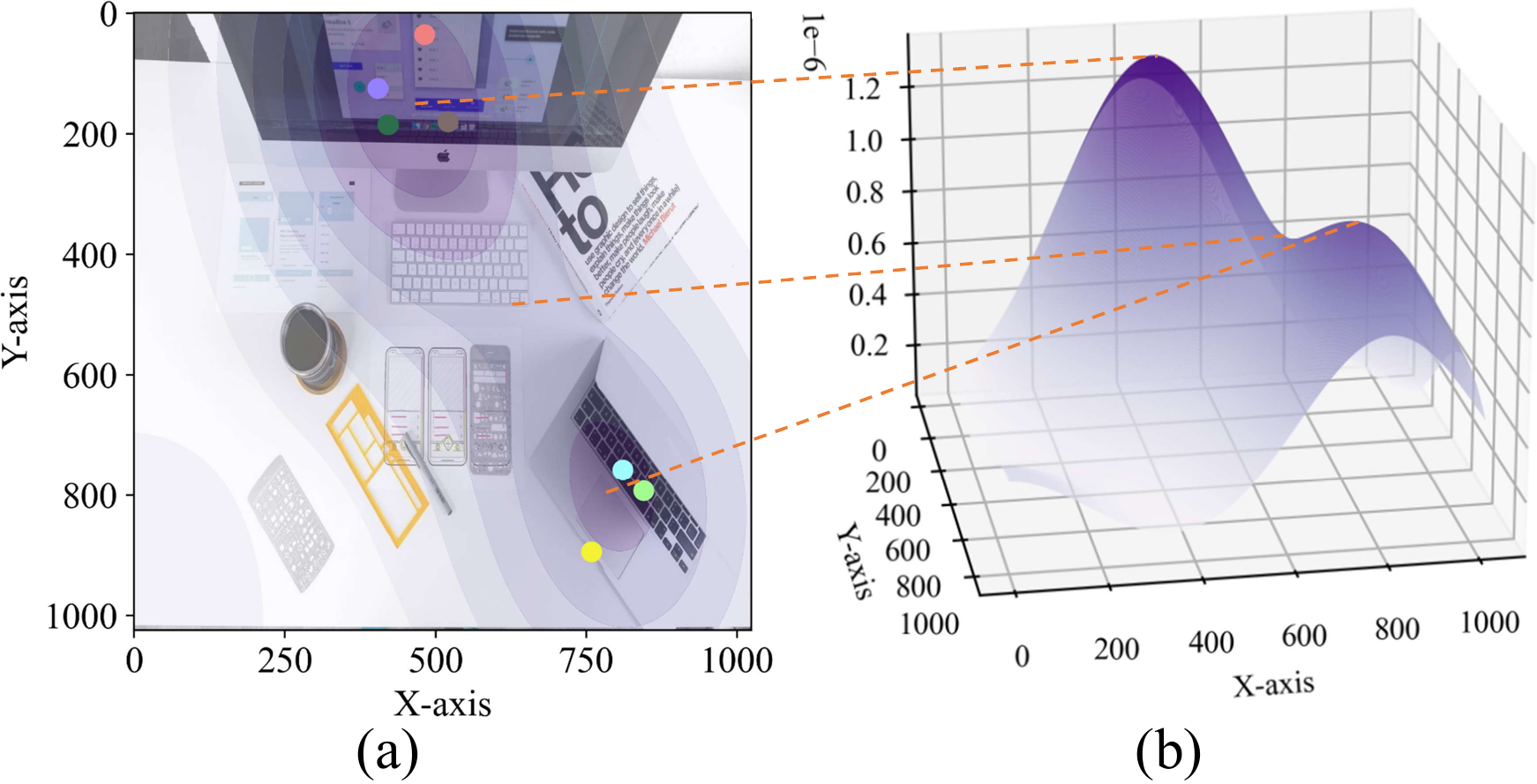}   
  \caption{Illustration of our probability density score. (a) An example image with the gaze points of multiple subjects when they heard ``the computer''. (b) Its corresponding ground-truth distribution $\hat{P}_i$ visualized in a 3D way.}
  \label{pds}
  \vspace{-1em}
\end{figure}
 
\subsection{Probability Density Loss}\label{section:score_function}
The high individual variability of ground-truth gaze trajectories raises a challenge to optimization and evaluation. As the example in Figure~\ref{pds} (a) shows, when hearing ``the computer'', the gaze points of subjects stay on computers. However, there are two computers in the image, resulting in the gaze points being split into two groups. Such diverse targets make the commonly used Mean Square Error (MSE) loss confused. In the example, a middle point between two groups minimizes the MSE loss, which is undesired as no computer is there. Therefore, we need better objectives to measure whether a gaze trajectory is human-like.

We propose a distribution-based measure, called Probability Density Score (PDS), instead of point-based measures like Euclidean Distance. First, we estimate the distribution $\hat{P_i}$ formed by multiple ground-truth gaze points (using Gaussian kernel density estimation) as the ground-truth distribution for word $w_i$, as shown in Figure~\ref{pds} (b). Second, for a predicted gaze point $\hat{s}_i$, we measure how well it fits the ground-truth distribution by the normalized value of its probability density on this distribution:
\begin{equation}
    \textit{PDS}(\hat{s}_i)=\frac{\hat{P}_i(\hat{s}_i)}{\max_{s} \hat{P}_i(s)}.
\end{equation}
In the example, a point near to the center of any group can get higher score than the middle point between groups. For a predicted gaze trajectory, we average the scores for all gaze points within the trajectory and get the final trajectory PDS. 
We finally utilize negative trajectory PDS as the training loss, i.e., probability density
 (PD) loss: $L(\hat{S}) = - \sum^n_{i=1} \textit{PDS}(\hat{s}_i)$. PDS can also be used as a measurement, which shows excellent discriminative power in our comparison experiments.
\section{Experiments}

\begin{figure*}
  \centering
  \includegraphics[width=0.97\textwidth]{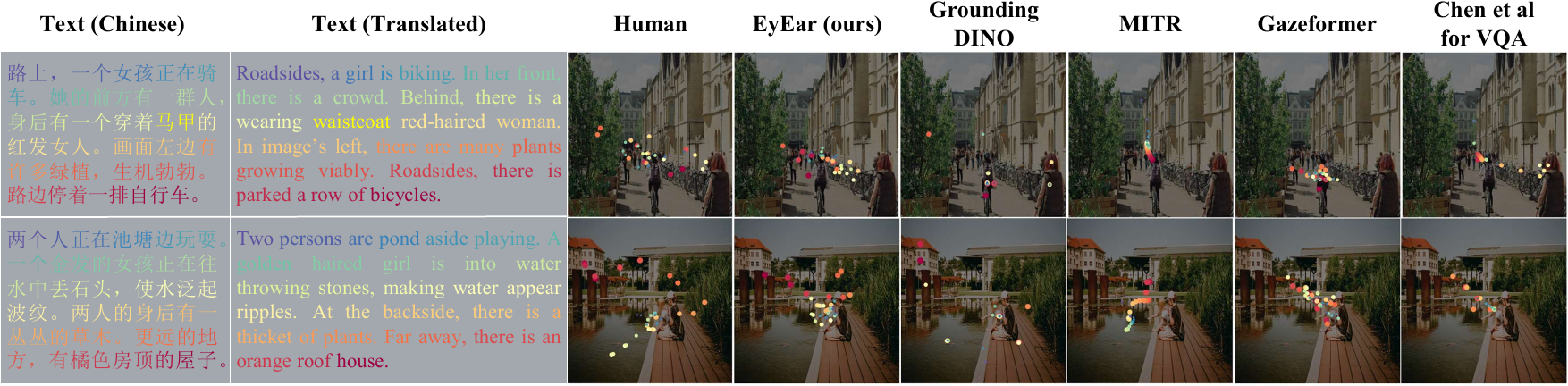} 
  \caption{Visualization of the predicted gaze trajectories of different models and the ground-truth human gaze trajectories. Best viewed in color. 
  We provide \textit{word-to-word} translations for better understanding.}
  \label{overall_result_fig}
  \vspace{-1em}
\end{figure*}

\subsection{Experiment Setup}

\textbf{Implementation Details.}
~~We randomly split our collected data into training, validation, and test sets in an 8:1:1 ratio.
We find that it is slow to optimize the PD loss, although it is more precise. Therefore, we adopt a two-stage training process for efficiency. In the first stage, we use MSE loss to optimize the model. After reaching a plateau in the loss, we continue training using our PD loss. All components of the model are jointly trained with the learning rate of 1e-4 and the optimizer is set to AdamW. Teacher forcing is utilized during training for sequence prediction. For more details, please refer to the Supplementary~B. \footnote{Our code and data are available at \url{https://github.com/XiaochuanLiu-ruc/EyEar}.}

\noindent\textbf{Baselines.}
~~Since our proposed task is new, there is no baseline model addressing this exact task. However, there are three types of methods that can be applied to solve our problem.
(1) \textbf{Pre-trained image-text models.} Our task can be considered as a downstream task of text-visual alignment, so we choose BriVL~\cite{huo2021-wenlan} and CLIP~\cite{radford2021-learning} as baselines.
(2) \textbf{Visual grounding models.} Visual grounding, which aims at precisely locating objects queried by natural language, can be seen as a highly simplified version of our task. We choose Grounding DINO (G-DINO)~\cite{liu2023-grounding} and MITR~\cite{meng2021-connectinga} as baselines.
(3) \textbf{Gaze trajectory prediction models.} Some gaze trajectory prediction models in other task scenarios, such as visual searching (VS) and visual question answering (VQA), can also be applied to our task, because they have text-image multimodal inputs. We choose recent high-performer works Chen et al for VS~\cite{chen2021predicting}, Gazeformer~\cite{mondal2023gazeformer} and Chen et al for VQA~\cite{chen2021predicting} as baselines.
We provide the detailed explanation of how each baseline is constructed in Supplementary~C.
In addition, following previous research, we also use human gaze trajectories to compare with each other to obtain \textbf{Human} inter-subject similarity as the upper bound of model performance.

\noindent\textbf{Metrics.}
~~In previous works, almost all widely-used metrics are \textbf{point-based metrics.} They treat a gaze trajectory as a sequence of gaze points and compare the similarity between the predicted and ground-truth sequences. Euclidean Distance (ED), Dynamic Time Wraping (DTW)~\cite{muller2007dynamic}, and ScanMatch~\cite{cristino2010-scanmatch} are widely used metrics. DTW uses a dynamic programming algorithm to discover an optimal alignment between two sequences, while ScanMatch uses the Needleman-Wunsch algorithm~\cite{needleman1970-general}. For point-based metrics, we average the scores compared to 8 ground-truth gaze trajectories as the final score.
Our proposed PDS is a \textbf{distribution-based metric}. PDS uses distribution fitted from multiple ground-truth gaze points to evaluate the predicted gaze points. 

\setlength{\tabcolsep}{0.8pt}
\begin{table}[t]
    \centering
    \begin{tabular}{lcccc}
    \toprule
        \textbf{Model} & \textbf{ED $\downarrow$} & \textbf{DTW $\downarrow$} & 
        \textbf{SMatch $\uparrow$} & 
        \textbf{PDS $\uparrow$}
        \\ \midrule
        \multicolumn{5}{l}{\textit{Pre-trained image-text models}} \\
        \textbf{CLIP} & 457.8 & 432.9 & 0.200 & 0.1689   \\
        \textbf{BriVL} & 437.6 & 402.6 & 0.239 & 0.1758  \\  
        \midrule
         \multicolumn{5}{l}{\textit{Visual grounding models}} \\
        \textbf{G-DINO} & 267.0 & 229.3 & \underline{0.443} & 0.4628 \\ 
        \textbf{MITR} & \underline{239.3} & 234.8 & 0.403 & 0.4792 \\ 
        \midrule
         \multicolumn{5}{l}{\textit{Gaze trajectory prediction models}} \\
        \textbf{Chen et al (VS)} & 443.3 & 407.8 & 0.253 & 0.1586   \\
        \textbf{Gazeformer} & 250.5 & 244.9 & 0.391 & 0.4807  \\
        \textbf{Chen et al (VQA)} & 249.4 & \underline{228.6} & 0.431 & \underline{0.5325} \\
         \midrule
        \multirow{2}{*}{\textbf{\modelname~(Ours)}} & \textbf{221.6}  & \textbf{201.3}  & \textbf{0.464} & \textbf{0.6138} \\
          & \textbf{(+8.0\%)}  & \textbf{(+11.9\%)} & \textbf{(+4.7\%)} & \textbf{(+15.3\%)}  \\        
        \midrule
        \textbf{Human} & 272.9 & 238.9 & 0.438 & 0.7243  \\ \bottomrule
    \end{tabular}
    \caption{Performance of different models on our test set. The best and runner-up are in \textbf{bold} and \underline{underlined}. Improvements are calculated between the best to the runner-up.}
    \label{overall_result}
    \vspace{-1em}
\end{table}

\subsection{Main Results}\label{experimental_result}

We compare~\modelname~with the baselines and present quantitative results in Table~\ref{overall_result}. It shows that~\modelname~consistently outperforms all the baselines in all metrics. The improvements over the best baselines are from 4\% to 15\% in different metrics. It indicates that although these baselines can be applied to our task, our method can better simulate human gaze. 

Our proposed PDS is the most reasonable measurement and has the best discriminative power among the four metrics. In terms of the metrics ED, DTW, and ScanMatch, our model and some baselines even outperform Human result, which is theoretically the upper bound of model performance. However, in our proposed PDS, there is still more than ten points room to be onpar with Human. This indicates that those point-based metrics are ineffective in distinguishing between natural and artificial gaze trajectories. Furthermore, all the metrics except for PDS yield close results for different models. In contrast, the evaluation results of PDS clearly differentiate the performance of various models. 

Among the three types of baselines, gaze trajectory prediction models perform the best, while pre-trained image-text models perform the worst. This is because the scenario of gaze trajectory prediction models is closer to our task. Chen et al for VQA performs the best among the baselines, which we attribute to the fact that the VQA gaze trajectory prediction task is structurally the closest to ours, as both involve images and sentences as input. However,~\modelname~significantly outperforms Chen et al for VQA by 15\% in terms of PDS. The improvements are attributed to the proposed three modules, which is indicated by ablation studies.

\subsection{Ablation Study}

\setlength{\tabcolsep}{5pt}
\begin{table}[t]
    \centering
    \begin{tabular}{lcccc}
    \toprule
        \textbf{Model} & \textbf{ED $\downarrow$} & \textbf{DTW $\downarrow$} & \textbf{SMatch $\uparrow$} & \textbf{PDS $\uparrow$} \\ \midrule
        \modelname & \textbf{221.6} & \textbf{201.3} & \textbf{0.464} & \textbf{0.614} \\
         \quad w/o Salience & 224.4 & 202.5 & 0.460 & 0.602 \\
         \quad w/o DynS & 231.3 & 207.3 & 0.460 & 0.568  \\ 
         \quad w/o GRU & 230.4 & 210.1 & 0.452 & 0.555 \\ 
         \quad w/o PD loss & 237.3 & 237.3 & 0.397 & 0.515  \\  
        \bottomrule
    \end{tabular}
    \caption{Overall ablation results. The best results are in \textbf{bold}.}
    \label{ablation_table}
    \vspace{-1em}
\end{table}

\subsubsection{Overall Results.}
We verify the effectiveness of dynamical system (DynS), GRU in Module 2, and PD loss by ablating one of them at a time. 
When ablating the dynamical system, we simply leverage feedforward neural networks to integrate audio semantic attraction point and vision salient attraction point: $\hat{s}_i= \alpha \cdot MLP_1(s^{Sal}) + (1-\alpha ) \cdot MLP_2(s^{Sem}_i)$.
We also remove vision salient attraction point, denoted by Salience, from DynS inputs to observe its impact. Table~\ref{ablation_table} shows the results. We have some findings from the table.
(1) Ablating any component results in a performance drop over all the metrics. This indicates that all the proposed components contribute to the superior performance of~\modelname. (2) Removing the PD loss incurs the largest performance drop consistently over all the metrics. It is expected, because a good loss function is crucial for model optimization. The probability density can better capture the loss between artificial and natural gaze trajectories from the perspective of statistics, making it better fit our data with high individual variability. (3) Ablating DynS and GRU results in similar performance drops, suggesting that capturing the motion patterns of eyes and achieving fine-grained audio-visual alignment in calculating the audio semantic attraction force hold equal importance in our task. (4) The removal of Salience has the least effect, indicating that the gaze trajectory is mainly influenced by the auditory stimuli, rather than the image itself, in our audio-visual scenarios.

\subsubsection{The effect of DynS.}
\begin{figure}[t]
    \centering
        \includegraphics[width=\linewidth]{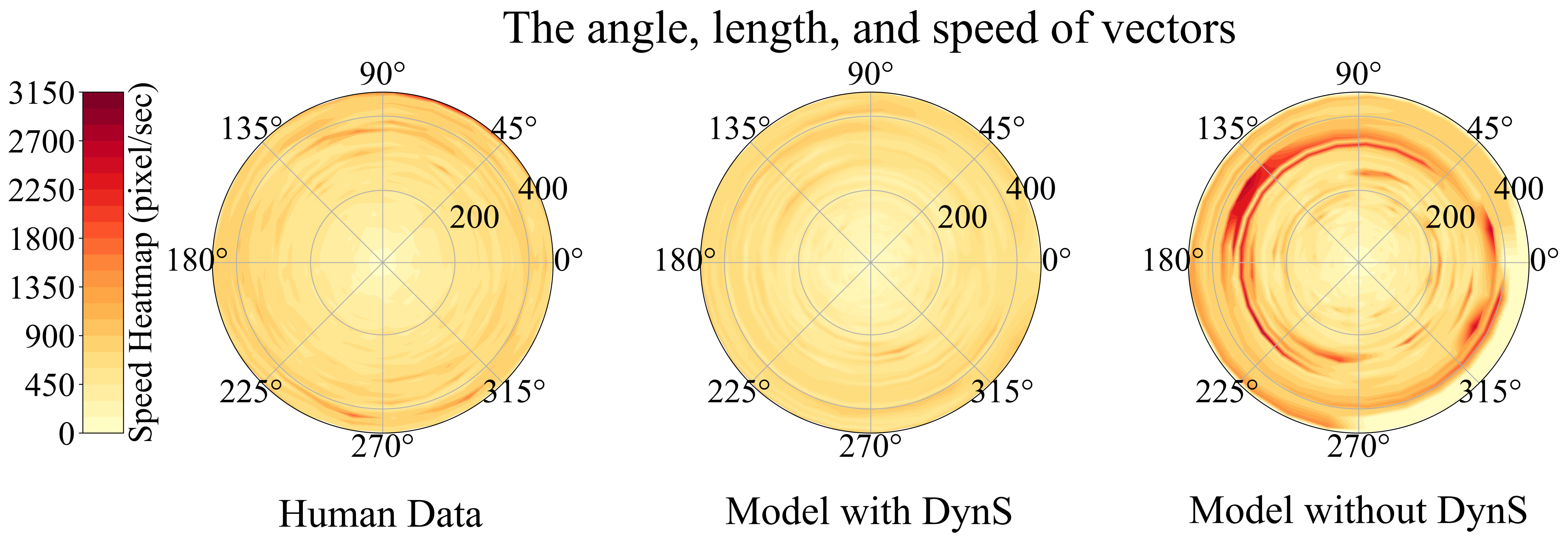}
        \caption{A radar chart showing the effect of DynS. Gaze trajectories are decomposed into saccade vectors pointing from the previous gaze point to the next. The degree of the polar coordinate represents the angle between the vectors and the horizontal direction. The radius of the polar coordinate is the length of the vectors. The heat map refers to the speed of vectors, calculated by length/duration.
        }
        \label{ablation_DS:sub2}
    \vspace{-1em}
\end{figure}

We further provide a visualization result to observe the effect of DynS. Similar to~\cite{dewhurst2012-it}, we first decompose the gaze trajectories into saccade vectors pointing from the previous gaze point to the next. Then, we statistically visualize the angle, length, and speed of these vectors, as shown in Figure~\ref{ablation_DS:sub2}. In these radar charts, each coordinate corresponds to a kind of saccade vectors, e.g., the coordinate (400, 45°) indicates the group of vectors with a length of 400 pixels and an angle of 45 degrees counterclockwise from the horizontal direction. The color intensity in charts indicates their average speed, where the darker red color represents higher speed. In human data, we can observe that the speed of saccade vectors remains stable as the length increases. When the length is the same, the speeds in different directions are similar to uniform. Our model with DynS also exhibits these phenomena; whereas the model without DynS exhibits much higher speeds (in darker red) for longer lengths. This indicates that DynS can learn the motion patterns of eyes and does not generate saccades with abnormal speed.

\subsection{Qualitative Analysis}

As shown in Figure~\ref{overall_result_fig}, we qualitatively compare the predicted gaze trajectories of different models to the ground-truth human gaze trajectories. It can be observed that~\modelname~predicts the most human-like gaze trajectories among all models in terms of both the fixation locations and the order of the fixations. \modelname~is able to not only gaze at the region when its corresponding object is mentioned in the audio, but also exhibit motion patterns similar to human eyes. For example, as shown in the last line of Figure~\ref{overall_result_fig}, the gaze trajectories of~\modelname~and human both make a loop and then move to the ``orange roof house", while other models either fail to gaze at the correct locations or fail to simulate the motion patterns of eyes. Moreover, the visual grounding models like Grounding DINO only predict separate points without movement process. However, we still have big room to improve because even our model cannot predict as large range motions as Human.

\section{Conclusion}
Aiming at enabling virtual characters to better mimic human eyes, we introduce a new task that aims to predict a gaze trajectory when a person views an image while hearing a narration. We collect a dataset with 20k gaze points to support related research. To address the challenges in our task, we propose a framework~\modelname. \modelname~uses a physics-informed dynamical system (DynS) to simulate the motion of eyes. In DynS, three potential forces affecting the eyes' motion are considered. The most important force comes from audio semantic attraction points, which we design a multimodal attention mechanism to predict. We also propose the probability density loss and score for better optimization and evaluation. \modelname~shows a notable performance gain, i.e., 15\% in probability density score (PDS), compared to the baselines, indicating the uniqueness and challenge of our task. However, there is still a gap between~\modelname~and human, leaving room for future improvements. In the future, we plan to transition from static images to more complex videos, aiming to restore the continuous changes in visual stimuli as experienced in the real world. We are also interested to experiment more open audio stimuli that is not narrative of an image. 

\section*{Ethical Statement}
Due to page limitation, the supplementary material and reproducing details are publically available at \url{https://github.com/XiaochuanLiu-ruc/EyEar/blob/main/AAAI2025_EyEar_Supplementary.pdf}.

\section*{Acknowledgements}
This work is supported by the National Natural Science Foundation of China (No. 62276268, No. 92270118).
We acknowledge Associate Professor Xiting Wang for providing valuable feedback and insightful suggestions that improved this paper. We would also like to express our gratitude to the Department of Psychology at Renmin University of China and Information Retrieval Lab at Tsinghua University for their support in providing the eye-tracking equipment used in this work.
\bibliography{main}

\begin{thebibliography}{40}
\providecommand{\natexlab}[1]{#1}

\bibitem[{Assens et~al.(2017)Assens, {Giro-i-Nieto}, McGuinness, and O'Connor}]{assens2017-saltinet}
Assens, M.; {Giro-i-Nieto}, X.; McGuinness, K.; and O'Connor, N.~E. 2017.
\newblock {{SaltiNet}}: {{Scan-Path Prediction}} on 360 {{Degree Images Using Saliency Volumes}}.
\newblock In \emph{2017 {{IEEE International Conference}} on {{Computer Vision Workshops}} ({{ICCVW}})}, 2331--2338.

\bibitem[{Birkhoff(1927)}]{birkhoff1927dynamical}
Birkhoff, G.~D. 1927.
\newblock \emph{Dynamical systems}, volume~9.
\newblock American Mathematical Soc.

\bibitem[{Borji and Itti(2015)}]{borji2015-cat2000b}
Borji, A.; and Itti, L. 2015.
\newblock {{CAT2000}}: {{A Large Scale Fixation Dataset}} for {{Boosting Saliency Research}}.
\newblock https://arxiv.org/abs/1505.03581v1.

\bibitem[{Chen et~al.(2020)Chen, Jiang, Yang, and Zhao}]{chen2020air}
Chen, S.; Jiang, M.; Yang, J.; and Zhao, Q. 2020.
\newblock Air: Attention with reasoning capability.
\newblock In \emph{Computer Vision--ECCV 2020: 16th European Conference, Glasgow, UK, August 23--28, 2020, Proceedings, Part I 16}, 91--107. Springer.

\bibitem[{Chen, Jiang, and Zhao(2021)}]{chen2021predicting}
Chen, X.; Jiang, M.; and Zhao, Q. 2021.
\newblock Predicting human scanpaths in visual question answering.
\newblock In \emph{Proceedings of the IEEE/CVF Conference on Computer Vision and Pattern Recognition}, 10876--10885.

\bibitem[{Chen et~al.(2021)Chen, Yang, Ahn, Samaras, Hoai, and Zelinsky}]{chen2021coco}
Chen, Y.; Yang, Z.; Ahn, S.; Samaras, D.; Hoai, M.; and Zelinsky, G. 2021.
\newblock Coco-search18 fixation dataset for predicting goal-directed attention control.
\newblock \emph{Scientific reports}, 11(1): 8776.

\bibitem[{Chen et~al.(2022)Chen, Yang, Chakraborty, Mondal, Ahn, Samaras, Hoai, and Zelinsky}]{chen2022characterizing}
Chen, Y.; Yang, Z.; Chakraborty, S.; Mondal, S.; Ahn, S.; Samaras, D.; Hoai, M.; and Zelinsky, G. 2022.
\newblock Characterizing target-absent human attention.
\newblock In \emph{Proceedings of the IEEE/CVF Conference on Computer Vision and Pattern Recognition}, 5031--5040.

\bibitem[{Cristino et~al.(2010)Cristino, Math{\^o}t, Theeuwes, and Gilchrist}]{cristino2010-scanmatch}
Cristino, F.; Math{\^o}t, S.; Theeuwes, J.; and Gilchrist, I.~D. 2010.
\newblock {{ScanMatch}}: A Novel Method for Comparing Fixation Sequences.
\newblock \emph{Behavior Research Methods}, 42(3): 692--700.

\bibitem[{Dewhurst et~al.(2012)Dewhurst, Nystr{\"o}m, Jarodzka, Foulsham, Johansson, and Holmqvist}]{dewhurst2012-it}
Dewhurst, R.; Nystr{\"o}m, M.; Jarodzka, H.; Foulsham, T.; Johansson, R.; and Holmqvist, K. 2012.
\newblock It Depends on How You Look at It: {{Scanpath}} Comparison in Multiple Dimensions with {{MultiMatch}}, a Vector-Based Approach.
\newblock \emph{Behavior Research Methods}, 44(4): 1079--1100.

\bibitem[{Engbert et~al.(2015)Engbert, Trukenbrod, Barthelm{\'e}, and Wichmann}]{engbert2015-spatial}
Engbert, R.; Trukenbrod, H.~A.; Barthelm{\'e}, S.; and Wichmann, F.~A. 2015.
\newblock Spatial Statistics and Attentional Dynamics in Scene Viewing.
\newblock \emph{Journal of Vision}, 15(1): 15.1.14.

\bibitem[{Huo et~al.(2021)Huo, Zhang, Liu, Lu, Gao, Yang, Wen, Zhang, Xu, Zheng, Xi, Yang, Hu, Zhao, Li, Zhao, Zhang, Song, Hong, Cui, Hou, Li, Li, Liu, Gong, Jin, Sun, Chen, Lu, Dou, Jin, Lan, Zhao, Song, and Wen}]{huo2021-wenlan}
Huo, Y.; Zhang, M.; Liu, G.; Lu, H.; Gao, Y.; Yang, G.; Wen, J.; Zhang, H.; Xu, B.; Zheng, W.; Xi, Z.; Yang, Y.; Hu, A.; Zhao, J.; Li, R.; Zhao, Y.; Zhang, L.; Song, Y.; Hong, X.; Cui, W.; Hou, D.; Li, Y.; Li, J.; Liu, P.; Gong, Z.; Jin, C.; Sun, Y.; Chen, S.; Lu, Z.; Dou, Z.; Jin, Q.; Lan, Y.; Zhao, W.~X.; Song, R.; and Wen, J.-R. 2021.
\newblock {{WenLan}}: {{Bridging Vision}} and {{Language}} by {{Large-Scale Multi-Modal Pre-Training}}.
\newblock arxiv:2103.06561.

\bibitem[{Jiang et~al.(2015)Jiang, Huang, Duan, and Zhao}]{jiang2015-dataset}
Jiang, M.; Huang, S.; Duan, J.; and Zhao, Q. 2015.
\newblock [{{DATASET}}] {{SALICON}}: {{Saliency}} in {{Context}}.
\newblock In \emph{2015 {{IEEE Conference}} on {{Computer Vision}} and {{Pattern Recognition}} ({{CVPR}})}, 1072--1080.

\bibitem[{Judd, Durand, and Torralba(2012)}]{judd2012-benchmark}
Judd, T.; Durand, F.; and Torralba, A. 2012.
\newblock A {{Benchmark}} of {{Computational Models}} of {{Saliency}} to {{Predict Human Fixations}}.

\bibitem[{Kamath et~al.(2021)Kamath, Singh, LeCun, Synnaeve, Misra, and Carion}]{kamath2021-mdetr}
Kamath, A.; Singh, M.; LeCun, Y.; Synnaeve, G.; Misra, I.; and Carion, N. 2021.
\newblock {{MDETR}} -- {{Modulated Detection}} for {{End-to-End Multi-Modal Understanding}}.
\newblock arxiv:2104.12763.

\bibitem[{Krauzlis, Lovejoy, and Z{\'e}non(2013)}]{krauzlis2013superior}
Krauzlis, R.~J.; Lovejoy, L.~P.; and Z{\'e}non, A. 2013.
\newblock Superior colliculus and visual spatial attention.
\newblock \emph{Annual review of neuroscience}, 36: 165--182.

\bibitem[{K{\"u}mmerer and Bethge(2021)}]{kummerer2021-stateoftheart}
K{\"u}mmerer, M.; and Bethge, M. 2021.
\newblock State-of-the-{{Art}} in {{Human Scanpath Prediction}}.
\newblock arxiv:2102.12239.

\bibitem[{K{\"u}mmerer, Bethge, and Wallis(2022)}]{kummerer2022-deepgaze}
K{\"u}mmerer, M.; Bethge, M.; and Wallis, T. S.~A. 2022.
\newblock {{DeepGaze III}}: {{Modeling}} Free-Viewing Human Scanpaths with Deep Learning.
\newblock \emph{Journal of Vision}, 22(5): 7.

\bibitem[{Lan, Scargill, and Gorlatova(2022)}]{lan2022-eyesyn}
Lan, G.; Scargill, T.; and Gorlatova, M. 2022.
\newblock {{EyeSyn}}: {{Psychology-inspired Eye Movement Synthesis}} for {{Gaze-based Activity Recognition}}.
\newblock In \emph{2022 21st {{ACM}}/{{IEEE International Conference}} on {{Information Processing}} in {{Sensor Networks}} ({{IPSN}})}, 233--246.

\bibitem[{Li et~al.(2022)Li, Zhang, Zhang, Yang, Li, Zhong, Wang, Yuan, Zhang, Hwang, Chang, and Gao}]{li2022-grounded}
Li, L.~H.; Zhang, P.; Zhang, H.; Yang, J.; Li, C.; Zhong, Y.; Wang, L.; Yuan, L.; Zhang, L.; Hwang, J.-N.; Chang, K.-W.; and Gao, J. 2022.
\newblock Grounded {{Language-Image Pre-training}}.
\newblock arxiv:2112.03857.

\bibitem[{Linardos et~al.(2021)Linardos, Kümmerer, Press, and Bethge}]{linardos2021deepgaze}
Linardos, A.; Kümmerer, M.; Press, O.; and Bethge, M. 2021.
\newblock DeepGaze IIE: Calibrated prediction in and out-of-domain for state-of-the-art saliency modeling.
\newblock arXiv:2105.12441.

\bibitem[{Liu et~al.(2023{\natexlab{a}})Liu, Li, Wu, and Lee}]{liu2023-visual}
Liu, H.; Li, C.; Wu, Q.; and Lee, Y.~J. 2023{\natexlab{a}}.
\newblock Visual {{Instruction Tuning}}.
\newblock arxiv:2304.08485.

\bibitem[{Liu et~al.(2013)Liu, Xu, Huang, Li, Xu, and Lin}]{liu2013-semanticallybased}
Liu, H.; Xu, D.; Huang, Q.; Li, W.; Xu, M.; and Lin, S. 2013.
\newblock Semantically-{{Based Human Scanpath Estimation}} with {{HMMs}}.
\newblock In \emph{2013 {{IEEE International Conference}} on {{Computer Vision}}}, 3232--3239.

\bibitem[{Liu et~al.(2022)Liu, Liang, Li, Huang, Zhang, Su, Zhu, and Zhang}]{liu2022-dqdetr}
Liu, S.; Liang, Y.; Li, F.; Huang, S.; Zhang, H.; Su, H.; Zhu, J.; and Zhang, L. 2022.
\newblock {{DQ-DETR}}: {{Dual Query Detection Transformer}} for {{Phrase Extraction}} and {{Grounding}}.
\newblock arxiv:2211.15516.

\bibitem[{Liu et~al.(2023{\natexlab{b}})Liu, Zeng, Ren, Li, Zhang, Yang, Li, Yang, Su, Zhu, and Zhang}]{liu2023-grounding}
Liu, S.; Zeng, Z.; Ren, T.; Li, F.; Zhang, H.; Yang, J.; Li, C.; Yang, J.; Su, H.; Zhu, J.; and Zhang, L. 2023{\natexlab{b}}.
\newblock Grounding {{DINO}}: {{Marrying DINO}} with {{Grounded Pre-Training}} for {{Open-Set Object Detection}}.
\newblock arxiv:2303.05499.

\bibitem[{Meng et~al.(2021)Meng, Yu, Zhang, Berg, Damavandi, Singh, and Bearman}]{meng2021-connectinga}
Meng, Z.; Yu, L.; Zhang, N.; Berg, T.; Damavandi, B.; Singh, V.; and Bearman, A. 2021.
\newblock Connecting {{What}} to {{Say With Where}} to {{Look}} by {{Modeling Human Attention Traces}}.
\newblock In \emph{2021 {{IEEE}}/{{CVF Conference}} on {{Computer Vision}} and {{Pattern Recognition}} ({{CVPR}})}, 12674--12683.

\bibitem[{Mondal et~al.(2023)Mondal, Yang, Ahn, Samaras, Zelinsky, and Hoai}]{mondal2023gazeformer}
Mondal, S.; Yang, Z.; Ahn, S.; Samaras, D.; Zelinsky, G.; and Hoai, M. 2023.
\newblock Gazeformer: Scalable, effective and fast prediction of goal-directed human attention.
\newblock In \emph{Proceedings of the IEEE/CVF Conference on Computer Vision and Pattern Recognition}, 1441--1450.

\bibitem[{Moschovakis, Gregoriou, and Savaki(2001)}]{moschovakis2001functional}
Moschovakis, A.; Gregoriou, G.; and Savaki, H. 2001.
\newblock Functional imaging of the primate superior colliculus during saccades to visual targets.
\newblock \emph{Nature neuroscience}, 4(10): 1026--1031.

\bibitem[{M{\"u}ller(2007)}]{muller2007dynamic}
M{\"u}ller, M. 2007.
\newblock Dynamic time warping.
\newblock \emph{Information retrieval for music and motion}, 69--84.

\bibitem[{Needleman and Wunsch(1970)}]{needleman1970-general}
Needleman, S.~B.; and Wunsch, C.~D. 1970.
\newblock A General Method Applicable to the Search for Similarities in the Amino Acid Sequence of Two Proteins.
\newblock \emph{Journal of Molecular Biology}, 48(3): 443--453.

\bibitem[{{Pont-Tuset} et~al.(2020){Pont-Tuset}, Uijlings, Changpinyo, Soricut, and Ferrari}]{pont-tuset2020-connecting}
{Pont-Tuset}, J.; Uijlings, J.; Changpinyo, S.; Soricut, R.; and Ferrari, V. 2020.
\newblock Connecting {{Vision}} and {{Language}} with {{Localized Narratives}}.
\newblock In Vedaldi, A.; Bischof, H.; Brox, T.; and Frahm, J.-M., eds., \emph{Computer {{Vision}} -- {{ECCV}} 2020}, Lecture {{Notes}} in {{Computer Science}}, 647--664. Cham: Springer International Publishing.
\newblock ISBN 978-3-030-58558-7.

\bibitem[{Radford et~al.(2021)Radford, Kim, Hallacy, Ramesh, Goh, Agarwal, Sastry, Askell, Mishkin, Clark, Krueger, and Sutskever}]{radford2021-learning}
Radford, A.; Kim, J.~W.; Hallacy, C.; Ramesh, A.; Goh, G.; Agarwal, S.; Sastry, G.; Askell, A.; Mishkin, P.; Clark, J.; Krueger, G.; and Sutskever, I. 2021.
\newblock Learning {{Transferable Visual Models From Natural Language Supervision}}.
\newblock arxiv:2103.00020.

\bibitem[{Shen and Zhao(2014)}]{shen2014-webpagea}
Shen, C.; and Zhao, Q. 2014.
\newblock Webpage {{Saliency}}.
\newblock In Fleet, D.; Pajdla, T.; Schiele, B.; and Tuytelaars, T., eds., \emph{Computer {{Vision}} -- {{ECCV}} 2014}, Lecture {{Notes}} in {{Computer Science}}, 33--46. Cham: Springer International Publishing.
\newblock ISBN 978-3-319-10584-0.

\bibitem[{Sun, Chen, and Wu(2021)}]{sun2021-visual}
Sun, W.; Chen, Z.; and Wu, F. 2021.
\newblock Visual {{Scanpath Prediction Using IOR-ROI Recurrent Mixture Density Network}}.
\newblock \emph{IEEE Transactions on Pattern Analysis and Machine Intelligence}, 43(6): 2101--2118.

\bibitem[{Tavakoli et~al.(2017)Tavakoli, Ahmed, Borji, and Laaksonen}]{tavakoli2017-dataset}
Tavakoli, H.~R.; Ahmed, F.; Borji, A.; and Laaksonen, J. 2017.
\newblock [{{DATASET}}] {{Saliency Revisited}}: {{Analysis}} of {{Mouse Movements Versus Fixations}}.
\newblock In \emph{2017 {{IEEE Conference}} on {{Computer Vision}} and {{Pattern Recognition}} ({{CVPR}})}, 6354--6362.

\bibitem[{Vaswani et~al.(2017)Vaswani, Shazeer, Parmar, Uszkoreit, Jones, Gomez, Kaiser, and Polosukhin}]{vaswani2017attention}
Vaswani, A.; Shazeer, N.; Parmar, N.; Uszkoreit, J.; Jones, L.; Gomez, A.~N.; Kaiser, {\L}.; and Polosukhin, I. 2017.
\newblock Attention is all you need.
\newblock \emph{Advances in neural information processing systems}, 30.

\bibitem[{Xia et~al.(2019)Xia, Han, Qi, and Shi}]{xia2019-predicting}
Xia, C.; Han, J.; Qi, F.; and Shi, G. 2019.
\newblock Predicting {{Human Saccadic Scanpaths Based}} on {{Iterative Representation Learning}}.
\newblock \emph{IEEE Transactions on Image Processing}, 28(7): 3502--3515.

\bibitem[{Xu et~al.(2014)Xu, Jiang, Wang, Kankanhalli, and Zhao}]{xu2014predicting}
Xu, J.; Jiang, M.; Wang, S.; Kankanhalli, M.~S.; and Zhao, Q. 2014.
\newblock Predicting human gaze beyond pixels.
\newblock \emph{Journal of vision}, 14(1): 28--28.

\bibitem[{Yang et~al.(2020)Yang, Huang, Chen, Wei, Ahn, Zelinsky, Samaras, and Hoai}]{yang2020-predicting}
Yang, Z.; Huang, L.; Chen, Y.; Wei, Z.; Ahn, S.; Zelinsky, G.; Samaras, D.; and Hoai, M. 2020.
\newblock Predicting {{Goal-Directed Human Attention Using Inverse Reinforcement Learning}}.
\newblock In \emph{2020 {{IEEE}}/{{CVF Conference}} on {{Computer Vision}} and {{Pattern Recognition}} ({{CVPR}})}, 190--199.

\bibitem[{Zanca, Melacci, and Gori(2020)}]{zanca2020-gravitational}
Zanca, D.; Melacci, S.; and Gori, M. 2020.
\newblock Gravitational {{Laws}} of {{Focus}} of {{Attention}}.
\newblock \emph{IEEE transactions on pattern analysis and machine intelligence}, 42(12): 2983--2995.

\bibitem[{Zaremba, Sutskever, and Vinyals(2014)}]{zaremba2014recurrent}
Zaremba, W.; Sutskever, I.; and Vinyals, O. 2014.
\newblock Recurrent neural network regularization.
\newblock \emph{arXiv preprint arXiv:1409.2329}.

\end{thebibliography}

\end{document}